# Women are slightly more cooperative than men (in one-shot Prisoner's dilemma games played online)

Valerio Capraro

Middlesex University London. V.Capraro@mdx.ac.uk

## Abstract

Differences between men and women have intrigued generations of social scientists, who have found that the two sexes behave differently in settings requiring competition, risk taking, altruism, honesty, as well as many others. Yet, little is known about whether there are gender differences in cooperative behavior. Previous evidence is mixed and inconclusive. Here I shed light on this topic by analyzing the totality of studies that my research group has conducted since 2013. This is a dataset of 10,951 observations coming from 7,322 men and women living in the US, recruited through Amazon Mechanical Turk, and who passed four comprehension questions to make sure they understand the cooperation problem (a one-shot prisoner's dilemma). The analysis demonstrates that women are more cooperative than men. The effect size is small (about 4 percentage points, and this might explain why previous studies failed to detect it) but highly significant (p<.0001).

## Introduction

Differences between men and women have intrigued generations of social scientists. Men, for example, have been found to be more competitive (Niederle & Vesterlund, 2006) and more risk seeking than women (Eckel & Grossman, 2008); whereas women have been found to be more honest (Abeler, Nosenzo & Raymond, in press; Capraro, 2017) and more altruistic than men (Brañas-Garza, Capraro, Rascón-Ramírez, 2018), especially when acting intuitively (Rand et al, 2016). Gender differences have also been found in moral judgments, where women typically make more deontological judgments in emotional dilemmas than men do (Friesdorf, Conway & Gawronski, 2015). There are gender differences in expectations too. Women are expected to be communal and unselfish, while men are expected to be agentic and independent (Eagly, 1987). And it has been observed that these differential expectations can have negative consequences on how people of different sex are judged when making the same action. For example, Heilman and Chen (2005) found that women are punished more than men when they fail to act altruistically.

Despite all this research, little is known about gender differences in cooperative behavior. Cooperation, that is, paying a cost to give a greater benefit to another person (or a group of people), while expecting the other person (or group of people) doing the same, is considered by many to be one of the behaviors that fundamentally characterize us, humans, as a species (Boyd, Gintis, Bowles & Richerson, 2003; Fehr & Gächter, 2002; Perc et al, 2017; Rand & Nowak, 2013). Some psychologists and anthropologists have advanced the hypothesis that the psychological mechanism underlying cooperative behavior, *shared intentionality*, might be the single trait that makes humans uniquely humans, since it is possessed by (non-autistic) children but not by great apes (Tomasello et al, 2015). It is thus undebatable that understanding whether there are gender differences in cooperative behavior is of primary importance. The question is also theoretically non-trivial, since both women and men have been adopting cooperative strategies since our hunter-gatherer ancestors (e.g., allomothering, hunting).

Numerous studies have tried to answer this question. See Croson and Gneezy (2009) for an exhaustive list of references. In short, the evidence is mixed and inconclusive. The literature offers examples in which men cooperate more than women (e.g., Rapoport & Chammah, 1965), examples in which women cooperate more than men (e.g., Frank, Gilovich & Regan, 1993), and examples in which there are no gender differences (e.g., Sell, Griffith & Wilson, 1993).

However, that previous studies failed to consistently detect gender differences in cooperative behavior does not necessarily mean that there are no gender differences. An alternative could be that gender differences are small and cannot be detected using the limited power of classical experiments conducted in physical laboratories. To shed light on this topic, I thus analyze the totality of experiments conducted by my research group and which use one-shot anonymous Prisoner's dilemma games, the standard measure of cooperative behavior among strangers (Nowak, 2006). This dataset contains 10,951 choices made by 7,322 men and women living in the US at the time of the experiments. The large size makes this dataset an excellent test for the hypothesis that there might be small gender differences in cooperative behavior, which previous studies failed to detect because of insufficient power. The following analysis indeed demonstrates that women cooperate more than men (49.7% vs 45.2%). Not surprisingly, the effect is very small. However, it is highly significant ($p<.0001$).

## Method

I analyze the entire set of experiments conducted by my research group and using the one-shot Prisoner's dilemma game. In these Prisoner's dilemma games, two players are matched anonymously over the Internet and they are shown the same set of instructions. They are given a certain amount of money (between $0.10 and $0.50 in my experiments) and they are asked whether they want to transfer (part of) this endowment to the other player. Any amount transferred is multiplied by a factor larger than 1 (between 1.1 and 10 in my experiments) and earned by the other player. The fact that the multiplier is larger than 1 implies that the two players are better off if they both transfer the money to each other than if they both keep it. However, each player has an individual incentive to keep the money. Thus, the Prisoner's dilemma captures the essence of the social dilemma of cooperation (Nowak, 2006).

Since 2013, my research group has conducted 31 studies using one-shot Prisoner's dilemma games. Some use a binary Prisoner's dilemma, where players can either transfer the whole endowment, or nothing; others use a (quasi)-continuous Prisoner's dilemma, in which players can transfer any amount from 0 to the whole endowment. Thus, the dependent variable will be a (quasi)-continuous variable called *Cooperation* normalized such that it takes value 1 if a subject transferred the full endowment. The combined dataset contains 10,951 observations collected on Amazon Mechanical Turk (Paolacci & Chandler, 2014) from individuals living in the US at the time of the experiment. All subjects passed four comprehension questions. One question asked which choice maximizes the player's payoff; one question asked which choice maximizes the other player's payoff; one question asked which choice by the other player maximizes the first player's payoff; one question asked which choice by the other player maximizes the other player's payoff. Thus, the subjects that I am going to analyze have a clear and complete understanding of the social dilemma structure of the game.

## Results

As a first step, I analyze the set of unique observations. To build this dataset, I start from the 10,951 observations in the combined dataset, and I check for multiple IP addresses and multiple TurkIDs. In doing so, I find 7,322 "unique" observations (mean age = 32.09). For each multiple observation, I keep only the first one, as determined by the date of participation in the experiment. In this dataset, simple average comparison shows that women cooperate slightly more than men (49.7% vs 45.2%). See Figure 1.

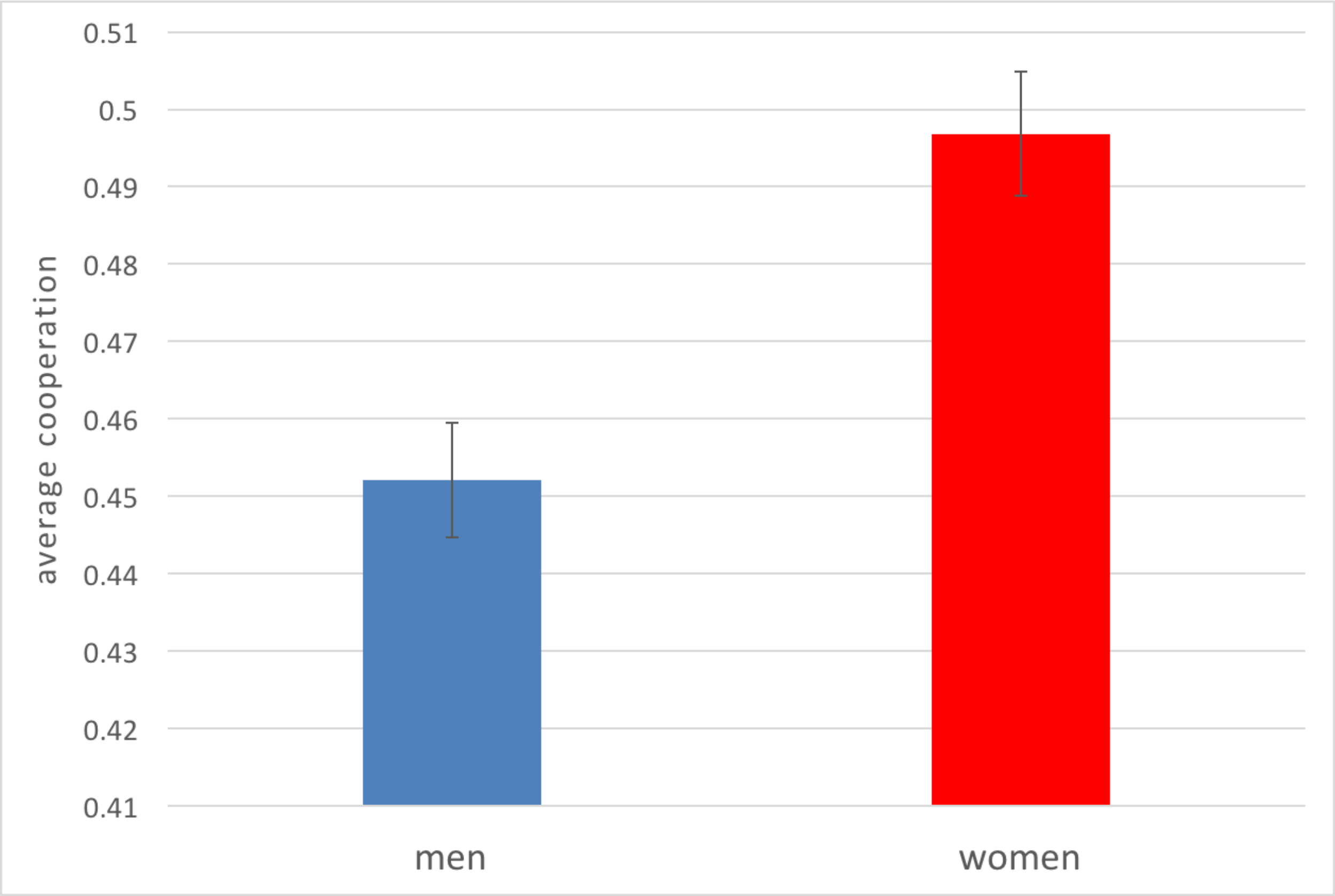

*Figure 1. Women are significantly more cooperative than men. N=7,322. Error bars = ± SEM*

Linear regression predicting *Cooperation* as a function of *Female* confirms the existence of a highly significant effect (coeff = 0.044, t=4.06, p<.0001). This effect is robust after controlling for age and education (coeff=0.0388, t=3.52, p<.0001).

As a second step, I conduct a meta-analysis of the 31 studies using the *metan* command in *Stata*. The forest plot is reported in Figure 2. There is a highly significant overall effect (overall effect size = 0.036, 95% CI [0.015, 0.057], z= 3.41, p=0.001), which is robust after controlling for age and level of education (overall effect size: 0.029, 95% CI [0.008, 0.050], z=2.73, p=0.006).

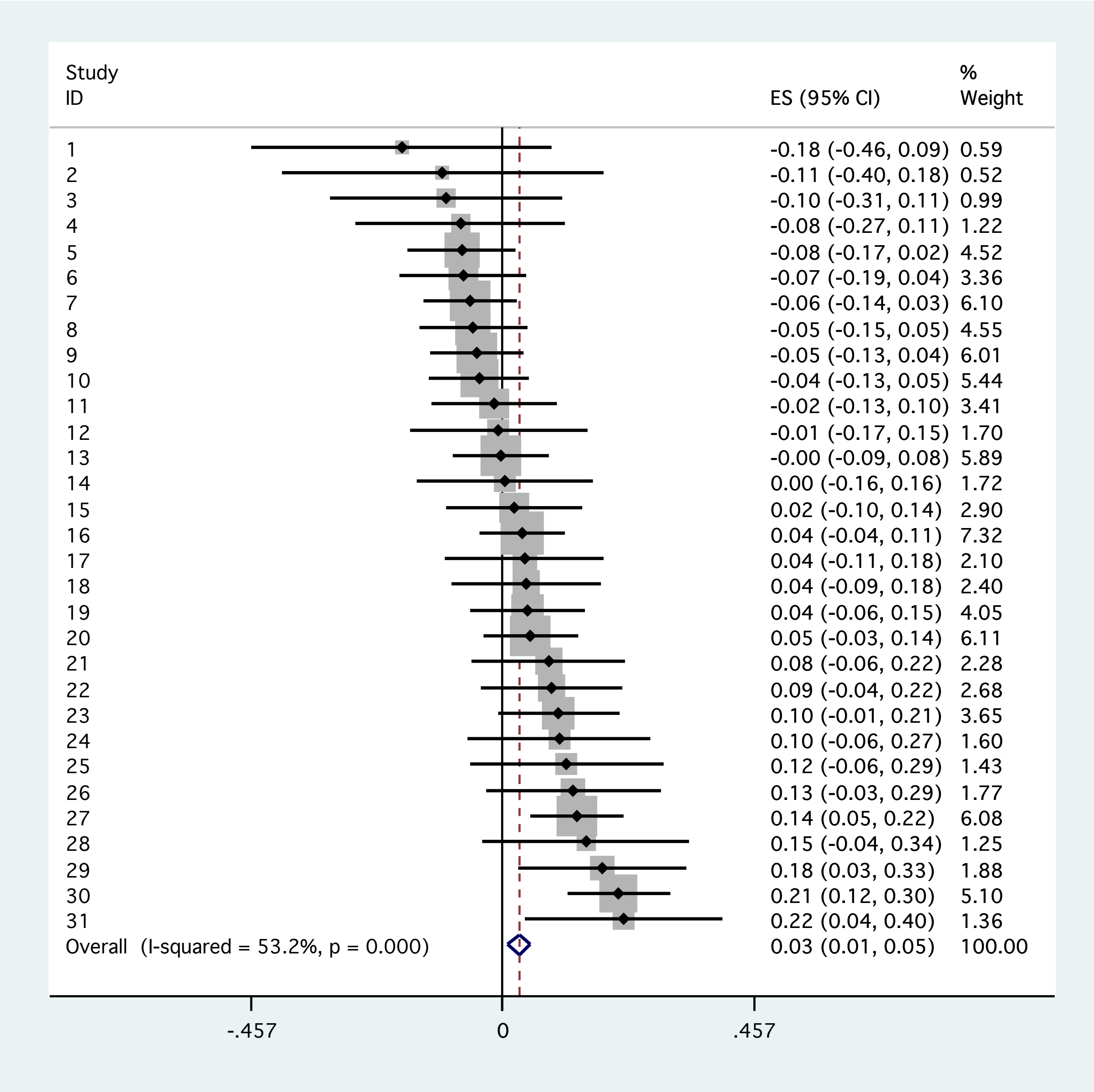

*Figure 2. Forest plot of the meta-analysis of the effects of sex on cooperative behavior on the 31 studies conducted by my research group. There is a significant overall effect such that women are more cooperative than men.*

I have also conducted a series of robustness checks. Instead of keeping, in case of multiple observations, only the first one, I have repeated the analysis by keeping the mean value of all choices from the same subject. Then I have repeated the analysis by keeping all observations, and treating them as independent. Finally, instead of splitting the dataset by "study", I repeated the meta-analysis by splitting the dataset by treatment. The result that women are significantly more cooperative than men is robust across all these specifications.

## Conclusion

Previous work on gender differences in cooperative behavior has been inconclusive. Here I shed light on this topic by analyzing a dataset of more than 10,000 observations on one-shot Prisoner's dilemma games played online by people living in the US. The analysis demonstrates that women are more cooperative than men. The effect is small (about 4 percentage points), and this might explain why previous studies failed to detect it. However, it is highly significant, and thus theoretically, if not practically, important.

These results have obvious limitations. They regard a specific population (MTurk workers living in the US) in a context in which cooperating is cheap (cost of cooperation between $0.10 and $0.50). Thus, one important direction for future research is to investigate the stability of gender differences in cooperative behavior across societies and for significantly larger stakes. Moreover, these results are silent regarding the theoretical mechanism underlying the revealed effect, and its potential practical consequences on real-life interactions. For example, are women punished more than men when failing to act cooperatively? These are important questions that need to be addressed in future research.